 \documentclass[nohyper,12pt,letterpaper]{JHEP3}
 \usepackage{graphicx}
 \usepackage{ytableau}
 \usepackage{amsmath}
 \usepackage{color}

 \def\bZ{{\mathbb Z}} 
 \def\bN{\mathbb{N}}

 \newcommand{\vev}[1]{{\left< {#1} \right>}}

\title{Exact results for Wilson loops in arbitrary representations}

\author{Bartomeu Fiol and Gen\'is Torrents  \\

Departament de F{\'\i}sica Fonamental i \\Institut de Ci{\`e}ncies del Cosmos, 

Universitat de Barcelona,

Mart{\'\i}\ i Franqu{\`e}s 1, 08028 Barcelona, Catalonia, Spain \\

\email{bfiol@ub.edu, genistv@icc.ub.edu }}

\abstract{We compute the exact vacuum expectation value of 1/2 BPS circular Wilson loops of ${\cal N}=4$ U(N) super Yang-Mills in arbitrary irreducible representations. By localization arguments, the computation reduces to evaluating  certain integrals in a Gaussian matrix model, which we do using the method of orthogonal polynomials. Our results are particularly simple for Wilson loops in antisymmetric representations; in this case, we observe that the final answers admit an expansion where the coefficients are positive integers, and can be written in terms of sums over skew Young diagrams. As an application of our results, we use them to discuss the exact Bremsstrahlung functions associated to the corresponding heavy probes.}  

\begin{document}
\section{Introduction}
Wilson loops are among the most interesting operators in any gauge theory. Their expectation values can serve as order parameters for the different phases of gauge theory. However, for generic four-dimensional gauge theories, the analytic evaluation of these expectation values for generic contours is currently out of reach. The situation improves for gauge theories with additional symmetries and Wilson loops with suitably chosen contours: for conformal theories with an AdS dual, it is possible to evaluate the vev of Wilson loops \cite{Rey:1998ik, Maldacena:1998im} with a variety of contours \cite{Berenstein:1998ij, Ishizeki:2011bf}, and in a variety of representations \cite{Drukker:2005kx, Yamaguchi:2006tq, Gomis:2006sb}. A second tool to compute vevs of Wilson loops is integrability, either of the dual string world-sheet \cite{Drukker:2005cu}, or of the planar limit of ${\cal N}=4$ SYM \cite{Drukker:2012de}. There have been also applications of the relations among certain 4d susy gauge theories and 2d CFTs \cite{Alday:2009aq} to the evaluation of Wilson loops \cite{Alday:2009fs}. Last but not least, it is also possible to use localization techniques to evaluate vevs of Wilson loops. In this regard one of the most remarkable results is due to Pestun \cite{Pestun:2007rz}, who showed that for ${\cal N}=2$ super Yang-Mills theories, localization arguments reduce the  evaluation of the expectation value of Euclidean half-BPS circular Wilson loops to a matrix model computation. For the particular case of ${\cal N}=4$ SYM, this had been anticipated in \cite{Erickson:2000af, Drukker:2000rr}. Localization techniques have also been applied to the evaluation of the vacuum expectation value of 't Hooft loops \cite{Gomis:2011pf}, loops preserving less supersymmetry \cite{Pestun:2009nn}  and 2-point functions of Wilson loops and local operators \cite{Giombi:2012ep}.  

The arguments in \cite{Pestun:2007rz} that reduce the evaluation of the vev of the Wilson loop to a matrix model computation are independent of the representation of the gauge group entering the definition of the Wilson loop. However, so far, the evaluation of the resulting matrix model integral has been carried out exactly only for a Wilson loop of ${\cal N}=4$ SYM in the fundamental representation \cite{Drukker:2000rr}, yielding a strikingly simple result in terms of a generalized Laguerre polynomial,
\begin{equation}
\vev{W_\Box(g)}=\frac{1}{N}L_{N-1}^1\left(-g \right)e^{\frac{g}{2}}
\label{exactfund}
\end{equation}
where we have defined $g=\lambda/4N$. The main goal of this work is to evaluate the relevant matrix model integrals to obtain the vev of half-BPS circular Wilson loops of ${\cal N}=4$ U(N) SYM in arbitrary irreducible representations. Half-BPS Wilson loops in higher rank representations have already been studied using holography by means of D-branes \cite{Drukker:2005kx, Yamaguchi:2006tq, Gomis:2006sb}. They have also been studied by solving the matrix model integrals in the large N limit for various representations \cite{Yamaguchi:2006tq, Hartnoll:2006is, Okuyama:2006jc} and also at strong coupling for arbitrary representations and gauge groups \cite{Gomis:2009ir}. Quite interestingly, this last reference managed to turn those results into non-trivial explicit checks of S-duality for ${\cal N}=4$ SYM.

Let us now explain what we have accomplished in the present work. The evaluation of the vev of the Wilson loop in the representation ${\cal R}$ of U(N) amounts to compute the vev of the insertion of $tr_{\cal R} e^X$ into the Gaussian Hermitian matrix model ($X$ is the Hermitian matrix being integrated over); after diagonalization in the matrix model, this insertion boils down to the Schur polynomial associated to the representation ${\cal R}$\footnote{In appendix B we have collected the definitions and some basic facts of the different basis of symmetric functions that appear in this work.}, as a function of the exponentials of the eigenvalues $x_i$,
$$
\vev{W_{\cal R}(g)}=\frac{1}{\dim \mathcal{R}}\vev{S_{\cal R}(e^{x_1},\dots,e^{x_N})}_{m.m.}
$$
Schur polynomials form a basis of the space of symmetric polynomials of $N$ variables, but as it turns out, they seem not to be the most convenient basis to carry out the integrals. Something similar was already encountered in the computation of the vev of Wilson loops in two-dimensional QCD \cite{Gross:1993yt}, where it was proposed to carry out the computations in the basis of multiply wound Wilson loops (which corresponds to the power sum symmetric functions, see appendix B), related to the Schur basis by Frobenius formula\footnote{See \cite{Gross:1998gk} for a similar discussion in the context of ${\cal N}=4$ SYM.}. Similarly, in this work, we manage to compute the exact vevs of insertions in the basis of monomial symmetric functions, from which one can recover the vevs in the Schur basis by a linear transformation. A couple of basics points of this linear transformation will help us to understand the structure of the answer we get for the vevs of Wilson loops. First, the Schur polynomial $s_{\cal R}$ can be written in terms of monomial polynomials $m_\tau$ with the same weight as ${\cal R}$  ({\it i.e.} the number of boxes of the associated Young diagrams is the same) and second, after imposing certain ordering among the partitions of $n$ (see appendix B), the matrix for this linear transformation is upper triangular when written using the ordering mentioned above, so only $\tau \leq {\cal R}$ representations have a non-zero contribution. The structure of the answer we get is then
\begin{equation}
\vev{W_{\cal R}(g)}=\frac{1}{\dim \mathcal{R}} \sum_{\tau \leq {\cal R}} K_{{\cal R} \tau} P_\tau(g) e^{\sum _i \tau_i^2\frac{g}{2}}
\label{genvev}
\end{equation}
where $K_{{\cal R} \tau}$ are positive integers (the {\it Kostka numbers}) that realize the linear transformation, and $P_\tau(g)$ are polynomials that we compute explicitly. Each polynomial multiplies an exponential,  with exponent given by the sum of the squares of the elements of the partition $\tau$, $\sum _i \tau_i^2$ (these exponentials were found already in \cite{Gomis:2009ir}).

The discussion above describes the generic case, but this general picture simplifies drastically for the particular case of Wilson loops in the antisymmetric representation of U(N). The reason is that Schur functions for the antisymmetric representation are already monomial symmetric functions, so in this case no change of basis is needed, and the final answer is given by a polynomial times an exponential, similar to the result found for the fundamental representation, eq. (\ref{exactfund}).       
We can write the resulting polynomial in a couple of ways: the matrix integral spits it out as a sum of products of generalized Laguerre polynomials, which is a straightforward but not terribly illuminating expression. We have managed to rewrite it as a single polynomial in $g$, and for the $k$-th antisymmetric representation we obtain a result of the form
\begin{equation}
\vev{W_{{\cal A}_k}(g)} = e^{\frac{g k}{2}}\; \sum_{j=0}^{k(N-k)} d_j(k,N)\frac{g^j}{j!} \hspace{1cm}d_j(k,N)\in \bN
\label{genantivev}
\end{equation}
where $d_j(k,N)$ are positive integers for which we derive a precise combinatorial formula, that we evaluate for specific values of $j$. An alternative and very succinct presentation of these results can be given in terms of the generating function of the vevs of these Wilson loops,
$$
\vev{F_A(t)}=\sum_{k=0}^Nt^{N-k}{N\choose k}\vev{W_{{\cal A}_k}}
$$
for which we obtain
$$
\vev{F_A(t)}=|t+A(g)e^{\frac{g}{2}}|
$$
where $A(g)$ is an $N\times N$ matrix with generalized Laguerre polynomials as entries, $A_{ij}=L_{i-1}^{j-i}(-g)$. The result for the fundamental representation,  eq. (\ref{exactfund}), follows by taking the trace of this matrix.

Besides its intrinsic interest, having explicit exact results for these Wilson loops can have a number of applications that we will discuss in the last section of the paper. In the present note, we will use these results to discuss the exact Bremsstrahlung functions for the corresponding heavy probes. For any heavy probe coupled to an arbitrary four-dimensional conformal field theory, the Bremsstrahlung function determines many relevant properties, like the total radiated power \cite{Correa:2012at, Fiol:2012sg} and the momentum diffusion coefficient \cite{Fiol:2013iaa} of an accelerated probe. For the specific case of 1/2 BPS heavy probes of ${\cal N}=4$ SYM it was shown in \cite{Correa:2012at} that the Bremsstrahlung function can be derived from the vev of the 1/2 BPS circular Wilson loop,
\begin{equation}
B(\lambda,N)_{\cal R}=\frac{1}{2\pi^2}\lambda\partial_\lambda \hbox{ log}\vev{W_{\cal R}}
\label{introbrem}
\end{equation}
so once we have these vevs in a given representation, it is straightforward to obtain the Bremsstrahlung function. It was observed in \cite{Fiol:2012sg, Fiol:2013iaa} that the Bremsstrahlung function for a heavy probe in the fundamental representation is, for fixed N, a rational function of the 't Hooft coupling, and it becomes linear both at weak and a strong coupling. 
In the light of the general structure presented in  (\ref{genvev}) for generic representations and in (\ref{genantivev}) for the antisymmetric representation, it follows from (\ref{introbrem}) that while the linearity in the 't Hooft coupling at weak and strong coupling (for fixed N) is common to all Bremsstrahlung functions, exact Bremsstrahlung functions are rational functions of the coupling only when there is a single exponential in the vev of the Wilson loop, {\it i.e.}, for antisymmetric representations.

The outline of the paper is as follows. In section 2 we start by  considering the vevs of Wilson loops in antisymmetric representations. We solve the matrix integral explicitly and discuss some properties of the generating function for these vevs. We point out that the resulting vevs admit a certain expansion with positive integer coefficients. The combinatorial formulas for these coefficients involve a number of ingredients suggestive of an interpretation in terms of fermions on a lattice, but we have not managed to come up with a satisfactory physical realization of these integers. In section 3 we turn to arbitrary representations; we first consider a perturbative expansion in $g$ for the matrix model integral, at finite N. We point out that this expansion involves evaluating {\it shifted Schur functions} \cite{okounkov}; we manage to compute at every order in $g$ the part of the coefficient that is of highest degree in the Casimir invariants of ${\cal R}$, and obtain a quite simple result involving the second Casimir of ${\cal R}$,
$$
\vev{W_{R}(g)}=\sum_{n=0}^\infty \left [\left(\frac{c_2({\cal R})}{2}\right)^n+\dots\right ] \frac{g^n}{n!}
$$
We then turn to computing the exact expectation value for a basis of symmetric functions, the monomial symmetric functions. In section 4 we discuss the Bremsstrahlung function for the corresponding heavy probes, using the results obtained in the previous sections. We conclude in the last section by briefly mentioning possible directions for future research. We have included three appendices: two with brief summaries on skew Young tableaux and on symmetric functions, and a third one with an alternative proof of one of the results in the main text.

\section{Wilson loop in antisymmetric representation}
In this work we are concerned with some specific non-local operators in ${\cal N}=4$ SYM. As argued in the seminal works  \cite{Rey:1998ik, Maldacena:1998im}, in ${\cal N}=4$ SYM it is natural to generalize the usual Wilson loop to include scalar fields. Locally BPS Wilson loops are then determined by a representation ${\cal R}$ and a contour ${\cal C}$ 
\begin{equation}
W_{\cal R}[{\cal C}]=\frac{1}{\hbox{dim }{\cal R}}\hbox{Tr}_{\mathcal R}{\cal P}\hbox{exp }
\left(i \int_{\cal C} (A_\mu \dot x^\mu +|\dot x|\Phi_i \theta^i)ds \right)
\end{equation}
We have fixed the overall normalization of the Wilson loop by the requirement that at weak coupling, $\vev{W_{\cal R}}=1+{\cal O}(g)$. We will restrict ourselves to half BPS Wilson loops. The simplest case is an infinite straight line, with an arbitrary representation. For any representation, this Wilson loop has a trivial vev, $\vev{W_{\cal R}}=1$, due to the large amount of supersymmetry. By a special conformal transformation, this Wilson loop can be mapped to a circular Wilson loop in Euclidean signature (or in Lorentzian signature, to a loop with hyperbolic contour \cite{Branding:2009fw, Fiol:2011zg}). This conformal mapping does not, however preserve the value of the vev. It was argued in  \cite{Erickson:2000af, Drukker:2000rr} and later proven in  \cite{Pestun:2007rz} that the vev of 1/2 BPS circular Wilson loops of ${\cal N}=4$ SYM can be computed exactly by means of a Gaussian Hermitian matrix model. After diagonalization, the partition function of this matrix model is given by
\begin{equation}
{\cal Z}= \int \prod_{i=1}^N \frac{dx_i}{2\pi} 
\prod_{i<j} |x_i-x_j|^2 e^{-\frac{1}{2g}\sum _{k=1}^N x_k^2}
\label{partfunc}
\end{equation}
The vev of the 1/2 BPS Wilson loop in an arbitrary representation ${\cal R}$ is given by the expected value of the Schur polynomial of ${\cal R}$ 
\begin{equation}
\vev{W_{\cal R}(g)}=\frac{1}{\hbox{dim }{\cal R}} \frac{1}{{\cal Z}}\int \prod_{i=1}^N \frac{dx_i}{2\pi} 
\prod_{i<j} |x_i-x_j|^2 S_{{\cal R}}\left(e^{x_1},\dots,e^{x_N}\right)
e^{-\frac{1}{2g}\sum _{k=1}^N x_k^2}
\end{equation}
In the next section we will discuss this integral for arbitrary irreducible representations ${\cal R}$. It turns out that the case of  antisymmetric representations is particularly simple and the results are most explicit, so we will start with it. For the k-antisymmetric representation, the Schur polynomial is given by the k-th {\it elementary symmetric function} $e_k$\footnote{This can be seen for instance from the dual Jacobi-Trudi identity, see appendix B.}, so
\begin{equation}
\vev{W_{{\cal A}_k}(g)}=\frac{1}{{\cal Z}}\int \prod_{i=1}^N \frac{dx_i}{2\pi} 
\prod_{i<j} |x_i-x_j|^2 e^{x_1+\dots+x_k} e^{-\frac{1}{2g}\sum _{k=1}^N x_k^2}
\label{wilsonmm}
\end{equation}
From this integral, eq. (\ref{wilsonmm}),  it is straightforward to relate the vevs of the Wilson loops for the k-th and the (N-k)-th antisymmetric representations. To do so,  complete the squares for the $x_1,\dots,x_k$ eigenvalues in (\ref{wilsonmm}), and then change variables $\tilde x_i=x_i-g$. Except for the $x_i$-independent exponents generated by completing squares, the resulting integral is the one that yields the vev of the Wilson loop in the $(N-k)$-th representation, so we arrive at the relation
\begin{equation}
\vev{W_{{\cal A}_k}(g)}e^{-\frac{kg}{2}}=\vev{W_{{\cal A}_{N-k}}(g)}e^{-\frac{(N-k)g}{2}}
\label{particlehole}
\end{equation}
For future reference, we define the following generating function for the elementary symmetric functions $e_k$\footnote{This generating function is closely related to the usual one (see appendix B), $E(t)=\sum_{k=0}^N e_k(y)t^k=\prod_{i=1}^N (1+y_it)$. Indeed, $F_A(t)=t^NE(1/t).$},
$$
F_A(t)=\sum_{k=0}^N e_kt^{N-k}=\prod_{i=1}^N\left(t+e^{x_i}\right)
$$
so its expectation value serves as the generating function for the expectation values of Wilson loops in antisymmetric representations,
\begin{equation}
\vev{F_A(t)}=\sum_{k=0}^Nt^{N-k}{N\choose k}\vev{W_{{\cal A}_k}(g)}
\end{equation}
To compute the integral (\ref{wilsonmm}) we will apply the method of orthogonal polynomials 
(see e.g. \cite{Marino:2004eq} for a recent introduction),  following and generalizing the approach in \cite{Drukker:2000rr}. In a nutshell, we introduce a family of polynomials $p_n(x)$ orthogonal with respect to the measure $dx \; e^{-\frac{x^2}{2g}}$. In the case at hand these polynomials are closely related to the Hermite polynomials
$$
p_n(x)\equiv \left(\frac{g}{2}\right)^{\frac{n}{2}}H_n\left(\frac{x}{\sqrt{2g}}\right)
$$
since
\begin{equation}
\int \frac{dx}{2\pi}\; p_m(x)p_n(x)\; e^{-\frac{x^2}{2g}}=
n!g^n\sqrt{\frac{g}{2\pi}}\delta_{mn}\equiv h_n \delta_{mn}
\label{orthopol}
\end{equation}
It is straightforward to prove that the partition function (\ref{partfunc}) admits a very simple expression
\begin{equation}
{\cal Z}=N! \prod_{k=0}^{N-1} h_k
\label{partfunch}
\end{equation}
The relevance of these orthogonal polynomials for the computation at hand becomes apparent when we realize that we can substitute the Vandermonde determinant in (\ref{wilsonmm}) by a determinant of orthogonal polynomials,
$$
\vev{W_{{\cal A}_k}(g)}=
\frac{1}{{\cal Z}}\int \prod_{i=1}^N \frac{dx_i}{2\pi} |p_{i-1}(x_j)|^2 e^{x_1+\dots+x_k} e^{-\frac{1}{2g}\sum _{k=1}^N x_k^2}
$$
We now expand the determinants of orthogonal polynomials in terms of permutations of its matrix elements
$$
|p_{i-1}(x_j)|^2=\sum _{\sigma_1 \in S_N} (-1)^{\epsilon(\sigma_1)} \prod _{k_1=1}^N p_{\sigma_1(k_1)-1}(x_{k_1})
\sum _{\sigma_2 \in S_N} (-1)^{\epsilon(\sigma_2)} \prod _{k_2=1}^N p_{\sigma_2(k_2)-1}(x_{k_2})
$$
The crucial point in the argument is that for the eigenvalues $x_{k+1},\dots,x_N$
the integrals in (\ref{wilsonmm}) are not modified by the insertion of the Wilson loop operator, so due to the orthogonality of the polynomials, for a given $\sigma_1\in S_N$, the only $\sigma_2$s that survive integration are those for which $\sigma_2(m)=\sigma_1(m)$ for $m>k$. This means that in order to contribute to the full matrix model integral, $\{\sigma_2(1),\dots,\sigma_2(k)\}$ must be a permutation of $\{\sigma_1(1),\dots,\sigma_1(k)\}$. Let's call this permutation $\mu$. The integral is now
$$
\vev{W_{{\cal A}_k}(g)}=
\frac{1}{N!}\sum _{\sigma \in S_N} \sum _{\mu \in S_k} (-1)^{\epsilon(\mu)} \int  \prod_{i=1}^k  \frac{dx_i}{2\pi} \frac{p_{\sigma(i)-1}(x_i) p_{\mu(\sigma (i))-1}(x_i)}{h_{\sigma(i)-1}} e^{-\frac{x_i^2}{2g}+x_i}
$$
where we already performed the integrals over the $x_{k+1},\dots,x_N$ eigenvalues using (\ref{orthopol}), and substituted the partition function ${\cal Z}$ using (\ref{partfunch}). The remaining integral can carried out explicitly \cite{gradshteyn} and it is given in term of generalized Laguerre polynomials,
\begin{equation}
L_n^\alpha(-g)=\sum_{j=0}^n {n+\alpha \choose n-j} \frac{g^j}{j!}
\label{thelaguerres}
\end{equation}
so we arrive at
\begin{equation}
\vev{W_{{\cal A}_k}(g)}=
\frac{e^{\frac{kg}{2}}}{N!}\sum _{\sigma \in S_N} \sum _{\mu \in S_k} (-1)^{\epsilon(\mu)}\prod_{m=1}^k
L_{\sigma(m)-1}^{\mu(\sigma(m))-\sigma(m)}(-g)
\label{antivev}
\end{equation}
As a check, if we set $k=1$, $\mu$ is the identity, and out of the $N!$ permutations in ${\cal S}_N$, $(N-1)!$ have the same $\sigma(1)$ so
$$
\vev{W_{{\cal A}_1}(g)}=\frac{e^{\frac{g}{2}}}{N!}(N-1)! \sum_{n=0}^{N-1}L_n^0(-g)=\frac{1}{N}L_{N-1}^1(-g) e^{\frac{g}{2}}
$$
recovering the result (\ref{exactfund}) of \cite{Drukker:2000rr}.
 
Before we proceed, it is important to notice that the result (\ref{antivev}) can be very succinctly stated in terms of the generating function for the antisymmetric representation. Define the $N\times N$ matrix 
$$
A(g)_{ij}= L_{i-1}^{j-i}(-g) 
$$
where $i,j=1,\dots,N$. The expression (\ref{antivev}) is then equivalent to
\begin{equation}
\vev{F_A(t)}=\sum_{k=0}^Nt^{N-k}{N\choose k}\vev{W_{{\cal A}_k}(g)}=\left |t+A(g)e^{\frac{g}{2}}\right |
\label{vevoffa}
\end{equation}
For notation purposes, it is very convenient to define the polynomial factor of $\vev{W_{{\cal A}_k}(g)}$ in (\ref{antivev}) as
\begin{equation}
\vev{W_{{\cal A}_k}(g)}=\frac{1}{{N\choose k}}P_k(g) e^{k\frac{g}{2}}
\label{defthep}
\end{equation}
The polynomials $P_k(g)$ have as generating function,
$$
\sum _{k=0}^NP_k(g)t^{N-k}=\left |t+A(g)\right |
$$
The contribution of the exponential factor in (\ref{vevoffa}) is easy to keep track of, and in what follows we will mostly discuss properties of $|t+A(g)|$ rather than $|t+A(g)e^{\frac{g}{2}}|$. A first property of $|t+A(g)|$ that is not manifest from the definition of $A$, is that it is a palindromic polynomial in $t$, that is,  $P_k(g)=P_{(N-k)}(g)$. This follows by construction from the relation (\ref{particlehole}), and can also be proven from the definition of the matrix $A$ (see appendix C). A second property of $|t+A(g)|$ is that  $P_k(g)$ can be written as a linear combination of the monomials $g^i/i!$ with positive integer coefficients. That is, we have
\begin{equation}
P_k(g)=\sum_{j=0}^{k(N-k)} d_j(k,N)\frac{g^j}{j!} \hspace{1cm}d_j(k,N)\in \bN
\label{definethed}
\end{equation}
It is easy to argue that the coefficients $d_j(k,N)$ are integers: the definition of the generalized Laguerre polynomials (\ref{thelaguerres}) implies that each entry in the matrix $A_{ij}(g)$ is a linear combination of terms $g^j/j!$ with integer coefficients, and in computing the determinant $|t+A|$, products, sums and subtractions of integers give integers; furthermore
$$
\frac{g^i}{i!}\frac{g^j}{j!}={i+j \choose i}\frac{g^{i+j}}{(i+j)!}
$$
so we can conclude that the coefficients of $|t+A(g)|$ are linear combinations of the monomials $g^i/i!$ with integer coefficients. The proof that these coefficients are all positive will take a little more effort, and we postpone it for a moment.

Although we won't dwell in this direction, it is possible to promote $t$ and $g$ to complex variables and consider $|t+A(g)|$ as a spectral curve \cite{Hartnoll:2006is}. The fact that $|t+A(g)|$ is a palindromic polynomial in $t$ implies that roots come in pairs $t_i, 1/t_i$ (except $t_i=-1$, that appears unpaired for $N$ odd)\footnote{If, in analogy with the analysis of the spectral curve of classical strings in $AdS_5\times S^5$ (see \cite{SchaferNameki:2010jy} for  a review), we define quasimomenta $p_j$ by $t_j=e^{ip_j}$, this $\bZ_2$ involution translates into the quasi-momenta coming in pairs $(p_j,-p_j)$.}. For $g$ real, if $t_i$ is a root, so is $t_i^*$. From the fact that all coefficients in $P_k(g)$  are real and positive, we learn that for $g>0$, the roots of  $|t+Ae^{g/2}|$ can't be positive real numbers. Numerical experimentation suggests the following picture: for $g>0$ all roots are real and negative; at $g=0$ all eigenvalues are equal to -1, and as $g\rightarrow +\infty$, half of them tend to $-\infty$ as powers of $g$, while  the other half are their pairs $1/t_i$ and tend to zero. This is consistent with the observation of \cite{Hartnoll:2006is} that at large N, the discrete zeros coalesce on a branch cut along the negative real axis.

It is easy to compute $|t+A(g)|$ at linear order in $g$. For $g=0$, the matrix $A$ is upper triangular, so the g-independent term in the determinant is immediately computed to be $(1+t)^N$. At linear order  in $g$, the matrix is no longer upper triangular, there are non-zero matrix elements immediately below the diagonal, $A_{i,i+1}=g$. It is nevertheless still straightforward to compute the determinant to this order by evaluating minors, and the final result is
$$
|t+A(g)|=(1+t)^N+{N \choose 2}t(t+1)^{N-2}g+{\cal O}(g^2)
$$
From this we deduce the first terms in the polynomial entering the vevs of the Wilson loops, and expanding the exponential, find the coefficient at order $g$,
\begin{equation}
\vev{W_{k,N}}=\frac{1}{{N\choose k}}
\left({N\choose k}+{N\choose 2}{N-2 \choose k-1}g+\dots\right)e^{\frac{gk}{2}}
=\left(1+\frac{c_2({\cal A}_k)}{2}g+\dots\right) 
\label{linearvev}
\end{equation}
In the next section we will prove that for an arbitrary representation ${\cal R}$, the term linear in $g$ has coefficient $c_2({\cal R})/2$.

We have derived a formula that gives the vevs of the Wilson loops in terms of a matrix that has generalized Laguerre polynomials as entries, eq. (\ref{antivev}). We are now going to derive a formula for the individual coefficients. In particular, we will recover the results in (\ref{linearvev}) and prove that the coefficients $d_j(k,N)$ in (\ref{definethed}) are positive.
Starting with equation (\ref{antivev}), the first observation is that for any given $\sigma \in S_N$, there are $(N-k)!$ permutations $\tilde \sigma\in S_N$ such that $\tilde \sigma (1)=\sigma(1),\dots, \tilde \sigma(k)=\sigma(k)$, and they all contribute the same to the sum in (\ref{antivev}). In fact, the same is true if $\{\tilde \sigma(1),\dots,\tilde \sigma(k)\}$ is a permutation of $\{\sigma(1),\dots,\sigma(k)\}$ so
\begin{align*}
P_k(g) & =& \sum_{0\leq \sigma_k<\dots <\sigma_1\leq N-1} & \sum _{\mu \in S_k} (-1)^{\epsilon(\mu)}\prod_{m=1}^k L_{\sigma(m)}^{\mu(\sigma(m))-\sigma(m)}(-g)= 
\\
&  = & \sum_{0\leq \sigma_k<\dots <\sigma_1\leq N-1} & \sum _{\mu \in S_k} (-1)^{\epsilon(\mu)}\prod_{m=1}^k \sum_{n_m=0}^{\sigma_m}{\sigma_{\mu(m)} \choose \sigma_m-n_m}\frac{g^m}{n_m!}
\end{align*}
To proceed, it is convenient to define $\tau_m=\sigma_m-n_m$. With the understanding that $\frac{1}{a!}=0$ for $-a\in \bN$ we can extend the range of $\tau_m$ to $N-1$
$$
P_k(g)=\sum_{0\leq \sigma_k<\dots <\sigma_1\leq N-1} \sum_{\tau_1,\dots, \tau_k=0}^{ N-1} 
\left(\prod_{n=1}^k \frac{g^{\sigma_n-\tau_n}}{(\sigma_n-\tau_n)!}\right)
\sum _{\mu \in S_k} (-1)^{\epsilon(\mu)}\prod_{m=1}^k {\sigma_{\mu(m)}\choose \tau_m}
$$
In the expression above, the last sum is antisymmetric in $\tau_m$, so we can restrict the sum over 
${\tau_m}$ to k-tuples of different $\tau_i$. We write it as a sum over ordered k-tuples and its permutations,
\begin{align*}
P_k(g )& = & \sum_{0\leq \sigma_k<\dots <\sigma_1\leq N-1\atop 0\leq \tau_k<\dots <\tau_1\leq N-1}
g^{\sum_m \sigma_m-\tau_m} & \sum_{\nu\in S_k} \prod_{m=1}^k\frac{1}{(\sigma_m-\tau_{\nu_m})!}
\sum_{\mu \in S_k} (-1)^{\epsilon(\mu)} \prod_{n=1}^k {\sigma_{\mu_n} \choose \tau_{\nu_n}} \\
 & = &\sum_{0\leq \sigma_k<\dots <\sigma_1\leq N-1 \atop 0\leq \tau_k<\dots <\tau_1\leq N-1}
g^{\sum_m \sigma_m-\tau_m} & \sum_{\nu\in S_k} \prod_{m=1}^k\frac{1}{(\sigma_m-\tau_{\nu_m})!}
\epsilon_{\nu_1 \dots \nu_k} \sum_{\mu \in S_k} (-1)^{\epsilon(\mu)} \prod_{n=1}^k {\sigma_{\mu_n}\choose \tau_n} \\
& = & \sum_{0\leq \sigma_k<\dots <\sigma_1\leq N-1 \atop 0\leq \tau_k<\dots <\tau_1\leq N-1}
g^{\sum_m \sigma_m-\tau_m} & \prod_{m=1}^k \frac{\tau_m!}{\sigma_m!}  \left | {\sigma_i \choose \tau_j} \right |^2
\end{align*}
where in the second line we used the properties of reordering the rows of a determinant. Collecting all the terms with $g^n$, we arrive then at a formula for the coefficients $d_n(k,N)$ in (\ref{definethed}) that makes manifest that they are positive,
\begin{equation}
d_n(k,N)=n! \sum_{0\leq \sigma_k<\dots \sigma_1\leq N-1 \atop 0\leq \tau_k<\dots \tau_1\leq N-1}
\prod_{m=1}^k \frac{\tau_m!}{\sigma_m!}  \left | {\sigma_i \choose \tau_j} \right |^2
\delta _{\sum (\sigma_m-\tau_m),n}
\label{bcoeff}
\end{equation}
As a check, if we set $k=1$ we arrive at
$$
d_n(1,N)= \sum_{\sigma=n}^{N-1}{\sigma \choose \sigma -n}= 
{N \choose n+1}
$$
reproducing the expansion of $P_1(g)=L_{N-1}^1(-g)$.

\subsection{Relation with skew Young tableaux and free fermions}
We have just derived an expression, eq. (\ref{bcoeff}), for the coefficients $d_n(k,N)$ of the polynomial $P_k(g)$ in (\ref{definethed}). We would like to recast it in terms of a sum over skew Young diagrams (see appendix A for a brief introduction). By pulling common factors out of the binomial determinants, we can write these coefficients in various ways,
\begin{align}
d_n(k,N)& =& n! 
\sum_{0\leq \sigma_k<\dots <\sigma_1\leq N-1\atop 0\leq \tau_k<\dots <\tau_1\leq N-1} &
\left(\prod_{m=1}^k \frac{\tau_m!}{\sigma_m!} \right) \left | {\sigma_i \choose \tau_j} \right |^2
\delta _{\sum_{m=1}^k (\sigma_m-\tau_m),n}= \nonumber \\
& =& n! \sum_{0\leq \sigma_k<\dots <\sigma_1\leq N-1 \atop 0\leq \tau_k<\dots <\tau_1\leq N-1} &
\left(\prod_{m=1}^k \frac{\sigma_m!}{\tau_m!} \right) \left | \frac{1}{(\sigma_i - \tau_j)!} \right |^2
\delta _{\sum_{m=1}^k (\sigma_m-\tau_m),n} = \nonumber \\
& =& n! \sum_{0\leq \sigma_k<\dots <\sigma_1\leq N-1 \atop 0\leq \tau_k<\dots <\tau_1\leq N-1} &
\left | {\sigma_i \choose \tau_j} \right |  \left | \frac{1}{(\sigma_i - \tau_j)!} \right | \delta _{\sum_{m=1}^k (\sigma_m-\tau_m),n}
\label{dcoeff}
\end{align}
An important observation is that the sum above can be restricted to pairs of k-tuples such that $\sigma_i\geq \tau_i$ for $i=1,\dots,k$. The reason is that if for some $j$ it happens that $\sigma_j<\tau_j$, the matrix with binomial coefficients in (\ref{bcoeff}) has a zero block in the upper right corner. The full determinant is then the product of determinants of the diagonal blocks, but the determinant of the lower diagonal block is zero, since it has a zero row. In what follows, it is understood that the sum in (\ref{dcoeff}) is restricted to $\sigma_i\geq \tau_i$.

Now, for every pair of k-tuples $\{\sigma\},\{\tau\}$, we define 
$$
\lambda_i=\sigma_i-k+i\hspace{1cm} \mu_i=\tau_i-k+i \hspace{1cm}i=1,\dots,k
$$
It is easy to see that $\lambda_i\geq \lambda_{i+1}$ and $\mu_i\geq \mu_{i+1}$, so $\{\lambda\}$ and $\{\mu\}$ are partitions. It also follows that $\lambda_i\leq N-k$ and $\mu_i\leq N-k$ so the corresponding Young diagrams can be both fitted in a rectangle with $(N-k)\times k$ boxes. If we denote by $L(m,n)$ be the set of all Young diagrams that fit into a board of $m$ rows and $n$ columns \cite{stanalgebra}, we have just argued that the partitions $\lambda$ and $\mu$ have Young diagrams in $L(k,N-k)$.

Finally, it is also easy to prove that $\lambda$ and $\mu$ are partitions satisfying $\lambda_i\geq \mu_i$, so the Young diagram of $\mu$ fits inside the Young diagram of $\lambda$, and it makes sense to consider its complement, the skew Young diagram $\lambda/\mu$, see table \ref{tableskew} in appendix A. The Kronecker delta in (\ref{dcoeff}) suggests that the coefficients $d_n(k,N)$ can be rewritten in terms of skew Young diagrams with $n$ boxes. An important step in this direction is to recognize the determinant in the second line of (\ref{dcoeff}) as the one that appears in the formula (\ref{aitkenform}) giving the number $f_{\lambda/\mu}$ of standard Young tableaux for the skew diagram $\lambda/\mu$,
\begin{equation}
d_n(k,N)= \frac{1}{n!} \sum_{\mu \subseteq \lambda \in L(k,N-k), \atop| \lambda|-|\mu|=n}
\prod_{m=1}^k \frac{(\lambda_m+k-m)!}{(\mu_m+k-m)!}   f_{\lambda/\mu}^2
\label{combid}
\end{equation}
The various expressions we have derived for the coefficients $d_n(k,N)$, either the original formulas (\ref{dcoeff}) or the one just derived, eq. (\ref{combid}), involve ingredients that have a number of combinatorial interpretations, and  in particular can be interpreted as counting paths of fermions in different lattices. The binomial determinant appearing in the first line of (\ref{combid}) was given a beautiful interpretation as counting the number of $k$ non-intersecting ({\it i.e.} fermionic) paths in a 2d-lattice, with the $k$-tuples $\{\sigma\}$ and $\{\tau \}$ giving respectively the initial and final conditions  for the $k$ paths \cite{gessel}. However, if we try to pursue this interpretation, we don't know what meaning we should assign to the prefactor in the first line of (\ref{combid}). 

A second possibility is to try to interpret these coefficients in terms of time-dependent processes for fermions on a 1d lattice. The first step in this direction is to map each Young diagram to a configuration of fermions (see {\it e.g.} \cite{zinn-justin}). A standard Young tableau of skew shape $\lambda/\mu$ is then interpreted as a time-dependent process, with $\mu$ being the initial configuration, $\lambda$ being the final configuration, and the labeling of the boxes indicating the order in which they appear \cite{zinn-justin, stanalgebra}. $f_{\lambda/\mu}$ counts the number of ways to evolve from $\mu$ to $\lambda$, but again we don't know of a clear combinatorial interpretation for the prefactor in (\ref{combid}).

Perhaps the best way of summarizing our lack of a simple combinatorial interpretation for these coefficients is the third line in (\ref{combid}); there the two determinants with interpretations outlined above appear, but as we have seen each one hints at a different physical realization.

\subsection{Evaluating the coefficients}
We have derived a formula for the coefficients $d_j(k,N)$, eq.  (\ref{bcoeff}). The index $j$ runs from $0$ to $k(N-k)$, and for arbitrary values of it, it seems doubtful that the sum can carried out explicitly. We will now evaluate these coefficients for a few values of $j$, close to the endpoints of its range. As a consistency check, all the explicit results we obtain satisfy $d_n(k,N)=d_n(N-k,N)$. An important ingredient in the evaluation is that as argued in the previous subsection, we can restrict to pairs of k-tuples such that $\sigma_i\geq \tau_i$ for $i=1,\dots,k$.

For $n=0$, both k-tuples have to be identical to contribute: $\sigma_i=\tau_i$. In this case the  matrix with entries ${\sigma _i\choose \tau_j}$ is lower triangular, the determinant in (\ref{bcoeff}) is 1, as well as the prefactor, so $d_0$ is given by the number of k-tuples,
$$
d_0={N\choose k}
$$
Alternatively, in the language of skew Young diagrams, $n=0$ corresponds to the case of $\lambda=\mu$ and $d_0$ is just counting the number of Young diagrams that fit into a rectangle with $(N-k)\times k$ boxes, which is precisely ${N \choose k}$ \ (proposition 6.3 in \cite{stanalgebra}).

For $n=1$, given a k-tuple $\tau_i$, the only k-tuples $\sigma_i$ that contribute are those where all the $\sigma_i=\tau_i$, except for precisely one element $\sigma_j=\tau_j+1$. For each of those cases the matrix with entries ${\sigma _i\choose \tau_j}$ is lower triangular, the determinant is $\sigma_j$ and the contribution in each case is $\sigma_j$. It remains to count how many such pairs of k-tuples there are, which is easily seen to be ${N-2 \choose k-1}$. Adding all contributions we obtain
$$
d_1={N\choose 2}{N-2\choose k-1}
$$
These two computations reproduce the result obtained for the Wilson loop by expanding $|t+A(g)|$
to linear order in $g$, eq. (\ref{linearvev}).

For $n=2$, there are two types of contributions. There are contributions from pairs of k-tuples when all $\sigma_i=\tau_i$ except for a single $\sigma_j=\tau_j+2$. There are also contributions from pairs of k-tuples when $\sigma_m=\tau_m$ except for two $\sigma$s, $\sigma_i=\tau_i+1$ and $\sigma_j=\tau_j+1$, with $i<j$. It is convenient to treat separately the cases where $\tau_j=\sigma_i$ (in which case the matrix fails to be lower triangular) and the case $\tau_j>\sigma_i$. By arguments very similar to the ones in the previous cases we arrive at
$$
d_2=\frac{N!}{12 (k-1)!(N-k-1)!}\left(3k(N-k)-N-1\right)
$$
This coefficient allows us to write the perturbative expansion of the antisymmetric Wilson loop to order $g^2$,
\begin{equation}
\vev{W_{{\cal A}_k}}=1+\frac{c_2({\cal A}_k)}{2}g+\left(\frac{1}{4}c_2({\cal A}_k)^2-\frac{N+1}{12}\left(c_2({\cal A}_k)-c_1({\cal A}_k) \right) \right) \frac{g^2}{2!}+\dots 
\label{antigtwo}
\end{equation}
Having computed the first three coefficients $d_j(k,N)$, we turn to the other end of the range, when $j$ is close to $k(N-k)$. For $n=k(N-k)$, there is only one term that contributes: $\sigma_i=N-k-1+i$, $\tau_j=j-1$. The determinant is 1, as can be proven by induction on $k$, for $N$ fixed. Therefore
$$
d_{k(N-k)}(k,N)= \left( k(N-k)\right)!\;  \frac{0! 1!\dots (k-1)!}{(N-1)! (N-2)!\dots (N-k)!}
$$

For $n=k(N-k)-1$, there are two terms that contribute. The first one has $\sigma_i=N-k-1+i$, $\tau_j=0,1,\dots,k-2,k$; the corresponding determinant is $N-k$. The second term has $\sigma_i=N-k-1,N-k+1,\dots,N-1$ and $\tau_j=j-1$; the corresponding determinant is $k$. Adding these two terms one obtains
$$
d_{k(N-k)-1}(k,N)= \left( k(N-k)\right)!\;  \frac{0! 1!\dots (k-1)!}{(N-1)! (N-2)!\dots (N-k)!}N
$$

\section{Arbitrary Representations}
In this section we will perform two different types of computations, both regarding Wilson loops in arbitrary representations ${\cal R}$ of the group U(N). First we will consider a perturbative expansion in $g$ for the vev of Wilson loops for arbitrary representation and at finite N. We will show that the Schur function evaluated on exponentials of eigenvalues admits an expansion in terms of Schur functions evaluated on eigenvalues, whose vevs are known exactly for the Gaussian Hermitian matrix model. The coefficients of this expansion turn out to be essentially {\it shifted Schur functions} \cite{okounkov}, evaluated on the components of the highest weight $\lambda$ of the representation ${\cal R}$. In order to bring the resulting expressions closer to familiar quantities, we would have to write these shifted Schur polynomials in terms of the Casimir invariants of the representation, $c_p({\cal R})$. This is quite easy at order $g$, as we show, but it already becomes quite cumbersome at higher orders. Nevertheless, if we settle for finding the highest degree Casimir combination appearing at each order $g^n$, this turns out to be a solvable problem with an extremely simple answer.

We then switch gears and compute the exact expectation value of a complete basis of the space of symmetric polynomials. The most convenient one turns out to be the basis of monomial symmetric functions; from these one can then recover the vevs of Wilson loops by a linear transformation.

\subsection{Perturbative computation} 
For  a generic representation ${\cal R}$, the vev of the Wilson loop is given by the following integral, 
\begin{equation}
\vev{W_{\cal R}(g)}=\frac{1}{\dim\mathcal{R}}\frac{1}{{\cal Z}}\int \prod_{i=1}^N \frac{dx_i}{2\pi} 
\prod_{i<j} |x_i-x_j|^2 S_{{\cal R}}\left(e^{x_1},\dots,e^{x_N}\right)
e^{-\frac{1}{2g}\sum _{k=1}^N x_k^2}
\end{equation}
We want to address the evaluation of this integral for arbitrary representation ${\cal R}$, for finite $N$ and perturbatively in $g$. By a rescaling of the eigenvalues $x_i\rightarrow \sqrt{g} x_i$ we learn that an expansion in $g$ amounts to expanding the exponentials of eigenvalues in the Schur function $S_{{\cal R}}\left(e^{x_1},\dots,e^{x_N}\right)$: to compute the term at order $g^n$ we need to expand the exponentials up to terms $x^{2n}_i$. 
The key ingredient to expand Schur functions around unity is the following formula (see section I.3, example 10 in \cite{macdonald}),
$$
s_\lambda(1+y_1,\dots,1+y_N)=\sum _{\mu} b_{\lambda _\mu} s_{\mu}(y_1,\dots,y_N)
$$
where $b_{\lambda \mu}$ is given by a determinant of binomial coefficients,
$$
b_{\lambda \mu}=\left | {\lambda_i+N-i \choose \mu_j+N-j} \right |
$$
Let's start by considering the computation to one-loop, {\it i.e.} to order $g$. At this stage, we prefer not to resort to some of the heavier machinery that we will introduce for higher loops, and keep the computation intuitive.  We need to expand the Schur function $S_{\cal R}(e^{x_1},\dots,e^{x_N})$ to order $x_i^2$, and the answer is

\ytableausetup{boxsize=6pt}
\begin{align}
S_{\cal R}(e^{x_1},\dots,e^{x_N})  & =  b_{{\cal R}\emptyset}+
b_{{\cal R}\; \ydiagram{1}}(x_1+\dots +x_N)+\left(\frac{1}{2}b_{{\cal R}\; \ydiagram{1}}+b_{{\cal R}\; \ydiagram{2}}\right)\left(x_1^2+\dots+x_N^2\right)+ \nonumber \\
& + \left(b_{{\cal R}\; \ydiagram{2}}+b_{{\cal R}\; \ydiagram{1,1}}\right)\sum_{i<j}x_i x_j+\dots
\label{expandschur}
\end{align}
The next step is to evaluate the binomial determinants $b_{{\cal R} \mu}$ that appear in (\ref{expandschur}). We will express the results in terms of the Casimir invariants of the representation ${\cal R}$, written as polynomials of the components of its highest weight $\lambda$. The generating function of the Casimirs of ${\cal R}$ can be written in terms of $\sigma_i=\lambda_i+N-i$ as \cite{barut}
\begin{equation}
G(z)=\sum c_p(\sigma)z^p=\frac{1}{z} \left[1-\prod_{i=1}^N\left(1-\frac{z}{1-\sigma_i z}\right)\right]
\label{genercasi}
\end{equation}
To this order only the first two Casimirs can appear,
$$
c_1({\cal R})=|\lambda|=\sum_i \lambda_i
$$
$$
c_2({\cal R})=\sum_i \lambda_i(\lambda_i+N+1-2i)
$$
We obtain
\begin{align}
b_{{\cal R}\emptyset}=\hbox{dim }{\cal R}, & & b_{{\cal R}\; \ydiagram{2}}=\frac{\hbox{dim }{\cal R}}{2n(n+1)}\left( c_2({\cal R})+c_1({\cal R})^2-(n+1)c_1({\cal R})\right) \nonumber \\
b_{{\cal R}\; \ydiagram{1}}=\hbox{dim }{\cal R}\frac{c_1({\cal R})}{N} & &
b_{{\cal R}\; \ydiagram{1,1}}=\frac{\hbox{dim }{\cal R}}{2n(n-1)}\left(- c_2({\cal R})+c_1({\cal R})^2+(n-1)c_1({\cal R})\right)
\label{thebdets}
\end{align}
The last ingredient we need are the vevs  $\vev{x_1+\dots+x_N}$, $\vev{x_1^2+\dots+x_N^2}$ and $\vev{\sum_{i<j} x_i x_j}$. These vevs can be easily computed with the method of orthogonal polynomials, making repeated use of the recurrence relations among the orthogonal polynomials \cite{Marino:2004eq}, but an ever easier way to compute them is to write them as linear combinations of the known n-point functions $\vev{\hbox{tr}X^2}$ and $\vev{\hbox{tr}X\hbox{tr}X}$. Either way, we obtain,
\begin{equation}
\vev{x_1+\dots+x_N}=0, \; \; \vev{x_1^2+\dots+x_N^2}=gN^2, \hspace{1cm} \vev{\sum_{i<j} x_i x_j}=-{N \choose 2}g
\label{easyvevs}
\end{equation}
Plugging the results for the coefficients (\ref{thebdets}) and the vevs (\ref{easyvevs}) back in the expansion (\ref{expandschur}), we finally obtain,
\begin{equation}
\vev{W_{\cal R}(g)}=1+\frac{c_2(\cal R)}{2}g+{\cal O}(g^2)
\label{genrepatg}
\end{equation}
In principle the method presented above can be used at higher loops. To compute efficiently the coefficients $b_{{\cal R} \mu}$ it is very useful to realize that they are essentially given by {\it shifted Schur functions}, $s_\mu^*({\cal R})$ \cite{okounkov}
\begin{equation}
b_{{\cal R} \mu}=\hbox{dim}{\cal R} \prod_{i=1}^N \frac{(N-i)!}{(\mu_i+N-i)!} s_\mu^*({\cal R})
\label{shiftedschur}
\end{equation}
The reason this connection is useful is that shifted Schur functions have many properties that generalize the properties of ordinary Schur functions; in particular, they admit a combinatorial definition in terms of reverse Young tableaux \cite{okounkov} which we have found quite efficient when it comes to actually computing them. We have carried out the computation of all necessary shifted Schur functions needed at order $g^2$ (i.e. corresponding to Young diagrams with up to 4 boxes). Unfortunately, the expressions obtained are long and far from illuminating, and we haven't found an efficient way to rewrite the coefficients $b_{{\cal R}\mu}$ in terms of the Casimir invariants that can appear: $c_1({\cal R}),\dots, c_4({\cal R})$.

In this work we will settle for a more modest problem, which is the following: assign a degree $p$ to the Casimir invariant $c_p({\cal R})$, corresponding to the highest power of $\lambda_i$ that appears in $c_p({\cal R})$, see (\ref{genercasi}). It follows that a product of Casimirs $c_{p_1} c_{p_2}$ has degree $p_1+p_2$, and it is easy to see that at order $g^n$ the highest degree that can appear is $2n$. For instance, at order $g^2$, the expansion of the Schur function $S_{\cal R}(e^{x_1},\dots,e^{x_N})$ will involve $c_4({\cal R}), c_3({\cal R})c_1({\cal R}),c_2({\cal R})^2,c_2({\cal R})c_1({\cal R})^2$ and $c_1({\cal R})^4$, plus terms of lower degree ({\it e.g.} $c_3({\cal R})$ or $c_1({\cal R})^2$). Our aim in what follows it to compute the term with highest degree possible ({\it i.e.} degree $2n$) in the coefficient of $g^n$, at every order $n$ in the $g$ expansion.

The first observation is that the only coefficients $b_{{\cal R} \mu}$ that contribute to the term with degree $2n$ are those with $\mu$ being a partition of $2n$: $\mu$ with a smaller number of boxes can only contribute to lower degree terms. By (\ref{shiftedschur}), this amounts to considering shifted Schur functions $s_\mu^*({\cal R})$,  with $\mu$ being a partition of $2n$. Furthermore, each $s_\mu^*({\cal R})$ can be written as a sum of the ordinary Schur function $s_\mu({\cal R})$ plus lower degree polynomials  \cite{okounkov} that don't contribute to the term we are considering.  To recapitulate, we have argued that the degree $2n$ term at order $g^n$ is given by
\begin{equation}
\hbox{dim}{\cal R} \sum_{|\mu|=2n} \prod_{i=1}^N \frac{(N-i)!}{(\mu_i+N-i)!} s_\mu({\cal R})
\vev{s_\mu(x)}
\label{sumone}
\end{equation}
In order to proceed, we will now make use that $\vev{s_\mu(x)}$ is known exactly for the Gaussian Hermitian matrix model \cite{DiFrancesco:1992cn}. The result of \cite{DiFrancesco:1992cn} can be written as
\begin{equation}
\vev{s_{\mu}(x)}=\frac{1}{2n!} \#[2^n] \chi_{\mu}[2^n] \prod_{i=1}^N\frac{(\mu_i+N-i)!}{(N-i)!}
\label{exactschur}
\end{equation}
where $[2^n]$ is the conjugacy class in ${\cal S}_{2n}$ with $n$ disjoint 2-cycles, and $\#[2^n]$ gives the number of elements in this conjugacy class, see eq. (\ref{elemconj}). When we plug (\ref{exactschur}) into (\ref{sumone}) the fractions of products cancel out. We can now write $s_\mu({\cal R})$ in terms of power sum polynomials, see eq. (\ref{frobenius}), 
$$
s_{\mu}({\cal R})=\frac{1}{2n!} \sum_\nu \#[\nu] \chi_\mu[\nu]p_\nu({\cal R})
$$
and make use of the orthogonality of the characters to simplify the coefficient (\ref{sumone}) as
$$
\hbox{dim }({\cal R})\frac{\#[2^n]}{(2n)!}p_{[2^n]}({\cal R})
$$
Now, $p_{[2^n]}({\cal R})$ differs from $c_2({\cal R})$ only in lower degree terms,  so for the purpose of computing the highest degree term, we can replace $p_{[2^n]}({\cal R})\rightarrow c_2({\cal R})$. Plugging the value for $\#[2^n]=(2n!)/(n! 2^n)$ - see eq. (\ref{elemconj}) - we arrive at the main result of this section,
$$
\vev{W_{R}}=\sum_{n=0}^\infty \left [\left(\frac{c_2({\cal R})}{2}\right)^n+\dots\right ] \frac{g^n}{n!}
$$
To reiterate, the dots stand for terms that we are missing at every order in $g^n$, that are of degree in Casimirs lower than $2n$, see (\ref{antigtwo}) for their explicit expression in the antisymmetric representation. These terms that we haven't computed come from different sources: first there are the contributions from shifted Schur functions with $|\mu|<n$; for instance, at order $g$ we see in (\ref{expandschur}) that $b_{{\cal R}\; \ydiagram{1}}$ also contributes, and in fact will contribute to every order. Second, when considering the shifted Schur functions with $|\mu|=n$, we replaced them by ordinary Schur functions, since they differ by lower degree polynomials. 

\subsection{Exact results in the monomial basis}
The Schur polynomials are the characters of irreducible representations of U(N), and they form a basis of the space of symmetric functions of $N$ variables, but it turns out that they don't form the most convenient basis to perform the integrals above. This is similar to what was encountered in the study of two-dimensional QCD \cite{Gross:1993yt}. The route taken there is to use Frobenius formula to relate the Schur polynomial to products of $\hbox {tr }U^k$ (i.e. multiply wound Wilson loops). In mathematical language this corresponds to changing basis from the Schur one to the {\it power sum symmetric functions}, $p_\lambda$. In the case at hand $\vev{p_k}$ is immediate to compute by a simple rescaling of the integral considered in \cite{Drukker:2000rr},
$$
\vev{p_k(g)}=\frac{1}{N}L_{N-1}^1(-k^2g)e^{k^2\frac{g}{2}}
$$
However, to evaluate the integral in a full basis of the ring of symmetric polynomials, we would have to compute $\vev{p_\lambda}$ for arbitrary partitions $\lambda$, and when facing this problem,  we end up having to compute $\vev{m_\lambda}$. For this reason, we consider the integrals above in the basis of {\it monomial symmetric polynomials} $m_\lambda$. These results can then  be used to write down the vevs of Schur polynomials, since they are related by a linear transformation $s_\lambda=\sum _\tau K_{\lambda \tau} m_\tau $, where the coefficients $K_{\lambda \tau}$ are positive integers, the so-called {\it Kostka numbers} (see appendix B).

The vev of monomial symmetric functions is 
\begin{equation}
\vev{m_\tau}=\frac{1}{{\cal Z}}\int \prod_{i=1}^N \frac{dx_i}{2\pi} 
\prod_{i<j} |x_i-x_j|^2 m_{{\tau}}\left(e^{x_1},\dots,e^{x_N}\right)
e^{-\frac{1}{2g}\sum _{k=1}^N x_k^2}
\end{equation}
since $m_\tau$ is a symmetric polynomial, and the rest of the integral is also symmetric, each monomial in $m_\tau$ gives the same contribution to the integral, so we can restrict to just one of them. Denote by $\ell(\tau)$ the number of non-zero entries of $\tau$, and by $\alpha_i(\tau)$ the number of times the number $i$ appears in the partition $\tau=(\tau_1,\tau_2,\dots)$. Then the number of monomials in  $m_\tau$ is 
$$
{N \choose \ell(\tau)} {\ell(\tau) \choose \alpha_1(\tau),\alpha_2(\tau),\dots}
$$
so
$$
\vev{m_\tau}={N \choose \ell(\tau)} {\ell(\tau) \choose \alpha_1(\tau),\alpha_2(\tau),\dots}
\frac{1}{{\cal Z}}\int \prod_{i=1}^N \frac{dx_i}{2\pi} \prod_{i<j} |x_i-x_j|^2 e^{\tau_1 x_1+\dots + \tau_{\ell(\tau)} x_{\ell(\tau)}}
e^{-\frac{1}{2g}\sum _{k=1}^N x_k^2}
$$
Now the argument proceeds along the same lines as in the previous section, for the antisymmetric representation. After substituting the Vandermonde determinant for a determinant of orthogonal polynomials, we expand these determinants. For the eigenvalues $x_{\ell(\tau)+1},\dots, x_N$ the insertion of the operator does not suppose any change, and we can carry out the integrals as always. The new integral that appears can again be computed explicitly \cite{gradshteyn} 
$$
\int \frac{dx}{2\pi} p_i(x)p_j(x) e^{\tau x} e^{-\frac{1}{2g}x^2}=
\sqrt{\frac{g}{2\pi}} g^j i! \tau^{j-i}L_i^{j-i}(-\tau^2g)e^{\frac{g}{2}\tau^2} 
$$
so we arrive at
\begin{equation}
\vev{m_\tau}=\frac{1}{N!}{N \choose \ell(\tau)} {\ell(\tau) \choose \alpha_1(\tau),\alpha_2(\tau),\dots}
\sum _{\sigma \in S_N} \sum _{\mu \in S_{\ell(\tau)}} (-1)^{\epsilon(\mu)} 
\prod_{m=1}^{\ell (\tau)} L_{\sigma(m)-1}^{\mu(\sigma(m))-\sigma(m)}(-\tau_m^2g) e^{\tau_m^2\frac{g}{2}}
\label{monovev}
\end{equation}
We can also introduce a generating functional for these vevs. It contains the antisymmetric representation as a particular case, but is more complicated. Define an infinite family of $N\times N$ matrices, labelled by $n\in N$
$$
A_{ij}^{(n)}(g)=n^{j-i}L_{i}^{j-i}(-n^2g)
$$
The vevs (\ref{monovev}) have then the following generating function
$$
\vev{\prod_{i=1}^N\left(t+\sum_{j=1}^\infty y_j e^{j x_i}\right)}=
\left |t+\sum_{j=1}^\infty y_j A^{(j)}(g) e^{j^2\frac{g}{2}}\right |
$$
Finally, as in the antisymmetric case, an explicit formula for the $\vev{m_{\tau}}$ can be found in terms of determinants of binomial coefficients:
\begin{equation}
\vev{m_{\tau }(g)} = \frac{e^{\frac{g}{2}\sum_{i} \tau_{i}^{2}} }{\prod_{i} \alpha_{i}!}  \sum_{c \in S_{\ell(\tau)}} \sum_{\mu \subseteq \lambda \in L(\ell,N-\ell)} \prod_{m=1}^\ell \frac{(\lambda_m+\ell-m)!}{(\mu_m+\ell-m)!} \left|\frac{\tau_{c_{i}}^{\lambda_{i}-\mu_{j}+j-i}}{(\lambda_{i}-\mu_{j}+j-i)!}\right|^2 g^{\sum \lambda_m-\mu_m}
\label{thebigsum}
\end{equation}
We have then managed to evaluate exactly the vevs of these symmetric functions in the Gaussian Hermitian matrix model. The vevs of the Schur polynomials are then linear combinations of these, with the transformation matrix given by {\it Kostka numbers}, see appendix B. The general form of the vevs of Wilson loops is then
\begin{equation}
\vev{W_{\cal R}(g)}=\frac{1}{\dim \mathcal{R}} \sum_{\tau \leq {\cal R}} K_{{\cal R} \tau} P_\tau(g) e^{\sum _i \tau_i^2\frac{g}{2}}
\label{generalvev}
\end{equation}
The polynomials in (\ref{generalvev}) can be read from either (\ref{monovev}) or (\ref{thebigsum}).
A relevant question, already answered in \cite{Gomis:2009ir}, is what is the largest exponent in (\ref{generalvev}), and it is immediate to see that it corresponds to $\sum _i \tau_i^2$ where $\tau$ is the highest weight of the representation ${\cal R}$.

\section{Bremsstrahlung functions}
Having devoted the previous sections to the computation of the exact vevs of half-BPS circular Wilson loops of ${\cal N}=4$ SYM, we would like to use the results obtained to discuss some properties of the associated heavy probes. We will do that by considering the corresponding Bremsstrahlung functions. The Bremsstrahlung function can be defined as the small angle limit of the cusp anomalous dimension, and for any heavy probe coupled to any four dimensional conformal field theory, it determines a number of interesting properties, like the two-point function of displacement operators \cite{Correa:2012at}, or the total radiated power \cite{Correa:2012at, Fiol:2012sg} and the momentum diffusion coefficient \cite{Fiol:2013iaa} of an accelerated probe.
Therefore computing the Bremsstrahlung function for probes of different conformal field theories is an interesting but in general difficult question. For $1/2$ BPS probes in ${\cal N}=4$ SYM, the situation is more favorable, since it was argued in \cite{Correa:2012at} that for these probes the Bremsstrahlung function can be derived from the vev of the corresponding circular Wilson loop
\begin{equation}
B_{\cal R}(\lambda,N)=\frac{1}{2\pi^2}\lambda\partial_\lambda \hbox{ log}\vev{W_{\cal R}}
\label{brefun}
\end{equation}
The argument leading to this relation is independent of the representation, so we can put to use our results for $\vev{W_{\cal R}}$. Since our results are most explicit for the antisymmetric representation, let's start with this case. For the antisymmetric representations, the vev of the Wilson loop is a polynomial in $g$ times an exponential, see eqs. (\ref{antivev}) or (\ref{defthep}). From this simple fact, it follows that when taking the logarithmic derivative in (\ref{brefun}) the final answer is a rational function in the coupling,
$$
B_{{\cal A}_k}^{U(N)}= \frac{\lambda}{16\pi^2 N}
\frac{\sum_{j=0}^{k(N-k)}
\frac{2d_{j+1}+kd_j}{j!}\left(\frac{\lambda}{4N}\right)^j}
{\sum_{j=0}^{k(N-k)}\frac{d_j}{j!} \left(\frac{\lambda}{4N}\right)^j}
$$
with the understanding that $d_{k(N-k)+1}=0$. For fixed $N$, both at weak and at strong 't Hooft coupling, the Bremsstrahlung function is linear in the 't Hooft coupling
$$
B_{{\cal A}_k}^{U(N)}= \frac{c_2({\cal A}_k) }{16\pi^2 N}\lambda \hspace{1cm} \lambda \ll 1
$$
$$
B_{{\cal A}_k}^{U(N)}= \frac{k}{16\pi^2 N} \lambda \hspace{1cm} \lambda \gg 1
$$
Let's now briefly discuss the case of general representations. Now $\vev{W_{\cal R}}$ is given by a linear combination of $\vev{m_\tau}$, so it is a sum of polynomials times exponentials, eq. (\ref{generalvev}). Since in general the exponents in these exponentials are different, it follows from (\ref{brefun}) that the corresponding Bremsstrahlung functions are no longer rational in the coupling. On the other hand, it also follows from (\ref{brefun}) that it is still true that for fixed $N$, both at weak and at strong 't Hooft coupling, the Bremsstrahlung function is linear in the 't Hooft coupling. The weak coupling result can be read off from (\ref{genrepatg}). In the large coupling limit, the coefficient of $\lambda$ is given by  the largest exponent in the exponentials which as pointed out after (\ref{generalvev}) (see also (\cite{Gomis:2009ir}) for a representation ${\cal R}$ with partition $\tau$ this exponent is $(\sum_i \tau_i^2)g/2$, so
$$
B_{{\cal R}}^{U(N)}= \frac{c_2({\cal R}) }{16\pi^2 N}\lambda \hspace{1cm} \lambda \ll 1
$$
$$
B_{{\cal R}}^{U(N)}= \frac{\sum_i \tau_i^2}{16\pi^2 N}\lambda \hspace{1cm} \lambda \gg 1
$$

\section{Outlook}
In closing, we would like to point out future directions and potential applications of the results obtained here.

A possible direction for further research is the use of our results as a concrete example of emergent spacetime out of a matrix model. Half-BPS Wilson loops have associated certain ten-dimensional solutions of IIB supergravity, that are given by a fibration over a Riemann surface \cite{D'Hoker:2007fq}. The relation between the classical geometry and the large N limit of the matrix model has been elucidated in \cite{Okuda:2008px}. It will be interesting to explore if having exact results for the vevs of these Wilson loops for any $N$ and $\lambda$ can lead to a deeper understanding of the emergence of these spacetimes.

Finally, in recent years there has been a fruitful interplay between matrix models and topological strings. For the model at hand, it was conjectured in \cite{Bonelli:2008rv} that antisymmetric Wilson loops of ${\cal N}=4$ are dual to amplitudes of the open topological string of Berkovits-Vafa \cite{Berkovits:2007rj}. It would be interesting to derive the results presented here by a direct computation of the relevant topological string amplitudes, and also to give an interpretation of positive integer coefficients $d_j$ in terms of enumerative geometry of the string background.

\section{Acknowledgements} 
We would like to thank Marcos Mari\~no for enjoyable conversations. BF would also like to thank the Theory Division at CERN for hospitality during part of this work. The research of BF is supported by MEC FPA2010-20807-C02-02, CPAN CSD2007-00042, within the Consolider-Ingenio2010 program, and AGAUR 2009SGR00168.  GT is supported by an FI scholarship by the Generalitat the Catalunya.

\appendix
\section{Skew Young tableaux}
In this appendix we recall a handful of definitions and formulas related to partitions and Young tableaux. The reader interested in more details is urged to consult any of the excellent books on this topic \cite{fulton, macdonald, stanenum}.

A {\it partition} is a sequence of non-negative integers in weakly decreasing order,
$$
\lambda_1\geq \lambda_2\geq \lambda_3\geq \dots
$$
The {\it weight} of the partition is $|\lambda|=\sum_i \lambda_i$. The {\it Young diagram} associated to a partition $\lambda$ is a collection of boxes arranged in left-justified rows of $\lambda_i$ boxes each. Given a partition $\lambda$, its {\it conjugate} partition $\lambda'$ is obtained by reflecting the Young diagram of $\lambda$ along its main diagonal, see table \ref{transposepart}.

\begin{table}
\centering
\ytableausetup{boxsize=normal}
\ytableausetup{nosmalltableaux}
\ydiagram{4,2,2} 
\hspace{1cm}
$\longrightarrow$
\hspace{1cm}
\ydiagram{3,3,1,1}
\caption{The conjugate partition $\lambda'$ is obtained by transposing the diagram of $\lambda$. In this example, $\lambda=(4,2,2)$ and $\lambda'=(3,3,1,1)$.}
\label{transposepart}
\end{table}

A {\it standard Young tableau} is a filling of a Young diagram of $|\lambda|=n$ boxes with the numbers $1,\dots,n$, such that the entries in each row and column are increasing. Given a partition $\{\lambda\}=\lambda_1\geq \dots\geq \lambda_k$, and its associated Young diagram, the number of standard Young tableaux with shape $\lambda$ can be written in various ways
\begin{equation}
f^\lambda=\frac{|\lambda|!}{\prod_x h(x)}=|\lambda|! \frac{\Delta(a)}{\prod _{m=1}^k a_m!}=
|\lambda|! \left|\frac{1}{(\lambda_i+j-i)!}\right|
\label{hooklength}
\end{equation}
where $a_i=\lambda_i+k-i$. The first formula is the famous hook length formula. The second one is proven in \cite{macdonald}. The third one is easy to prove starting from the second one.

Consider two partitions $\lambda,\mu$ such that $\mu_i\leq \lambda_i$. The {\it skew Young diagram} $\lambda/\mu$ is the complement of the diagram of $\mu$ inside the diagram of $\lambda$ (see table \ref{tableskew}).  The number of standard Young tableaux for a given skew shape $\lambda/\mu$ is given by the following formula, (corollary 7.16.3 in \cite{stanenum}) that generalizes the third one in (\ref{hooklength}),
\begin{equation}
f^{\lambda/\mu}=(|\lambda|-|\mu|)! \left|\frac{1}{(\lambda_i-\mu_j+j-i)!}\right|
\label{aitkenform}
\end{equation}

\bigskip
\begin{table}
\centering
\ytableausetup{boxsize=normal}
\ytableausetup{nosmalltableaux}
\ydiagram{4,2,2} 
\hspace{1cm}
\line(1,0){15}
\hspace{1cm}
\ydiagram{2,1}
\hspace{1cm}=\hspace{1cm}
\ydiagram{2+2,1+1,2}
\caption{The skew diagram $\lambda/\mu$ is the complement of $\mu$ in $\lambda$. In this example, $\lambda=(4,2,2)$ and $\mu=(2,1)$.}
\label{tableskew}
\end{table}

\section{Symmetric functions}
In this appendix we collect the definitions of the most common basis of symmetric polynomials in $N$ variables; they all happen to play a role in the main body of the paper. We also mention the change of basis relating Schur functions to the rest. These functions are labelled by partitions (see appendix A) that we denote by Greek letters, $\lambda,\mu,\dots$. Some excellent references on this topic are \cite{macdonald, fulton, stanenum}.

\medskip

\noindent
{\underline {\it Monomial symmetric functions.}} Given any partition $\lambda$ with at most $N$ non-zero elements, define the polynomials,
$$
m_\lambda(x_1,\dots,x_N)=\sum _\alpha x^\alpha
$$
where the sum is over distinct permutations $\alpha$ of $(\lambda_1,..)$.

\noindent
{\underline {\it Elementary symmetric functions}.}  For any integer $n\geq 0$ define
$$
e_n\equiv m_{1^n}
$$
and then for any partition $\lambda$ define
$$
e_\lambda=e_{\lambda_1} e_{\lambda_2}\dots
$$
The generating function for the polynomials $e_n(x_i)$ is
$$
E(t)=\sum_{k=0}^N e_k t^k = \prod_{i=1}^N (1+x_it).
$$

\bigskip

\noindent
{\underline {\it Complete symmetric functions}.} For any integer $n\geq 0$ define
$$
h_n=\sum_{|\lambda| = n}m_\lambda
$$
and then for any partition $\lambda$ define
$$
h_\lambda=h_{\lambda_1} h_{\lambda_2}\dots
$$
The generating function for the polynomials $h_n(x)$ is
$$
H(t)=\sum_{k=0}^N h_k t^k = \prod_{i=1}^N \frac{1}{(1-x_it)}=\frac{1}{E(-t)}
$$

\bigskip

\noindent
{\underline {\it Power sum symmetric functions}.} For any integer $n\geq 0$ define
$$
p_n\equiv m_n
$$
and then for any partition $\lambda$ define
$$
p_\lambda=p_{\lambda_1} p_{\lambda_2}\dots
$$
The generating function for the polynomials $p_n(x)$ is
$$
P(t)=\sum _{k\geq 1}p_kt^{k-1}=\sum_{i\geq i}\frac{x_i}{1-x_it}=
\frac{H'(t)}{H(t)}
$$

\medskip

\noindent
{\underline {\it Schur polynomials}.} For any partition $\lambda$, one possible way to define the Schur polynomials is as the quotient of two determinants,
$$
s_\lambda(x_1,\dots x_N)=\frac{|x_i^{\lambda_j+N-j}|}{|x_i^{N-j}|}
$$
There are linear relations that allow us to change basis among the five sets of polynomials just introduced. We will mention only the ones that relate Schur functions to the other ones. Schur polynomials can be written as determinants of matrices with either symmetric or elementary symmetric functions as entries,
$$
s_\lambda=|h_{\lambda_i-i+j}|=|e_{\lambda'_i-i+j}|
$$
These first equality is the Jacobi-Trudi identity, and the second one the dual Jacobi-Trudi identity ($\lambda'$ stands for the partition conjugate to $\lambda$, see appendix A). From the first one it follows that for the k-symmetric representation $s_{(k,0,\dots,0)}=h_k$, while from the second one it follows that for the k-antisymmetric representation, $s_{(1,\dots 1,0,\dots,0)}=e_k$.

Schur functions can be written in terms of monomial symmetric functions.
$$
s_\lambda=\sum_\tau K_{\lambda \tau} m_\tau
$$
The coefficients $K_{\lambda _\tau}$ are called {\it Kostka numbers}. They are positive integers, and they have a combinatorial meaning, they count the number of tableaux of shape $\lambda$ and weight $\tau$. Furthermore, let's introduce the {\it reverse lexicographic order} among all partitions of a given $n$: given two partitions of $n$, define this ordering as $\mu \leq \lambda$ if either $\mu=\lambda$ or the first non-zero $\lambda_i-\mu_i>0$. With this ordering, the Kostka matrix above is upper triangular, $K_{\lambda \tau}=0$ for $\lambda<\tau$.

Schur functions can also be written as linear combinations of power sum symmetric functions.
The relation involves some basics concepts of the representation theory of the symmetric group ${\cal S}_n$: the number of partitions of $n$ gives both the number of irreducible representations of ${\cal S}_n$ and the number of conjugacy classes, so each of these sets  can be labelled by these partitions. Any conjugacy class is characterized by the number of disjoint cycles of different length: the conjugacy class of ${\cal S}_n$, labelled by $\mu$, with permutations that are the product of $m_1$ cycles of length 1, $m_2$ of length 2 and so on, has a total number of elements given by
\begin{equation}
\#(\mu)=\frac{n!}{1^{m_1}m_1!2^{m_2}m_2!\dots}
\label{elemconj}
\end{equation}
Finally,  let's denote by  $\chi_\lambda$ the character of the irreducible representation labelled by $\lambda$. In terms of these ingredients, the relation between Schur and power sum polynomials is
\begin{equation}
s_\lambda= \frac{1}{n!} \sum_\mu \#(\mu) \chi_\lambda(\mu) p_\mu
\label{frobenius}
\end{equation}

\section{ $|t+A(g)|$ is a palindromic polynomial}
In this appendix we will present an alternative proof of  $|t+A(g)|$ being a palindromic polynomial, {\it i.e.} the fact that the coefficients $d_n(k,N)$ satisfy $d_n(k,N)=d_n(N-k,N)$. The idea of the proof is the following: the coefficient $d_n(k,N)$ is given by a sum of an expression evaluated over all skew Young diagrams $\lambda/\mu$ in $L(k,N-k)$ with $n$ boxes, 
\begin{equation}
d_n(k,N)= \frac{1}{n!} \sum_{\mu \subseteq \lambda \in L(k,N-k), \atop| \lambda|-|\mu|=n}
\prod_{m=1}^k \frac{(\lambda_m+k-m)!}{(\mu_m+k-m)!}   f_{\lambda/\mu}^2
\label{combidbis}
\end{equation}
For every skew Young diagram $\lambda/\mu$ we will first define partitions $\hat \mu'$ and $\hat \lambda'$ in $L(N-k,k)$ such that
$\hat \mu'/\hat \lambda'$ has also $n$ boxes, so there is a one-to-one correspondence among the diagrams contributing to $d_n(k,N)$ and those that contribute to $d_n(N-k,N)$. Finally we will show that both diagrams contribute the same to their respective coefficients, proving the equality.

Given a partition $\lambda=(\lambda_1,\dots, \lambda_k) \in L(k,N-k)$, define the complementary partition in $L(k,N-k)$ by
$$
\hat \lambda_i=N-k-\lambda_{k+1-i}
$$
$\hat \lambda$ is also a partition in $L(k,N-k)$. Graphically, it corresponds to taking the complement of the Young diagram of $\lambda$ in $L(k,N-k)$ and rotating it by 180$^\circ$ so it becomes an allowed Young diagram, see table \ref{hatlambdaprime}. Now consider $\hat \lambda '$, the partition conjugate to $\hat \lambda$. Graphically it corresponds to the transpose Young diagram. $\hat \lambda '$ is a partition in $L(N-k,k)$, see table \ref{hatlambdaprime}. Now for every skew Young diagram $\lambda/\mu$ contributing to $d_n(k,N)$ consider the skew Young diagram $\hat \mu'/\hat \lambda'$. This is a diagram in $L(N-k,k)$ with $n$ boxes, so it contributes to $d_n(N-k,N)$. We have established a one-to-one correspondence among the terms contributing to both coefficients. It remains to prove that these diagrams contribute the same.

\begin{table}
\centering
\ytableausetup{boxsize=normal}
\ytableausetup{nosmalltableaux}
\begin{multline}
\ydiagram[*(white)]{4+1,2+3,2+3}*[*(gray)]{5,5,5}
\hspace{.5cm} \Rightarrow \hspace{.5cm}
\ydiagram[*(white)]{3+2,3+2,1+4}*[*(gray)]{5,5,5}
\hspace{.5cm} \Rightarrow \hspace{.5cm}
\ytableaushort
{\none, \none, \none, \none, \none}
*{3,3,3,3,3}
*[*(gray)]{3,2,2,0,0}
\end{multline}
\caption{Starting with a partition $\lambda$ in the set $L(m,n)$, we define the partition $\hat \lambda$ by considering the complement of $\lambda$ in $L(m,n)$ and the rotating 180$^\circ$. We then take the transpose to obtain $\hat \lambda'$. In the example displayed $\lambda=(4,2,2)$ and $L(m,n)=L(3,5)$. Then $\hat \lambda=(3,3,1)$ and $\hat \lambda'=(3,2,2)$.}
\label{hatlambdaprime}
\end{table}

An important ingredient is the following result (see eq. (1.7) in  section I.1 of \cite{macdonald}). Consider a partition $\lambda \in L(k,N-k)$ and its transpose $\lambda'$. Define the numbers $a_i=\lambda+k-1$, $i=1,\dots,k$ and $c_j=k-1+j-\lambda'$, $j=1,\dots,N-k$. The numbers $a_i,c_j$ constitute a permutation of $(0,\dots,N-1)$.  For the partition $\hat \lambda'$ we have
$$
\hat a'_j=N-j-\lambda'_{N-k+1-j}=c_{N-k+1-j}
$$
and therefore by the previous result $a_i, \hat a'_j$ constitute a partition of $(0,\dots,N-1)$. The same is true for $b_i,\hat b'_j$ and from this it follows that
$$
\prod_{i=1}^k \frac{a_i!}{b_i!}=\prod_{j=1}^{N-k}\frac{\hat b_j'!}{\hat a'_j!}
$$
This shows that these factors in (\ref{combidbis}) are the same for $\lambda/\mu$ and $\hat \mu'/\hat \lambda'$. This together with $f_{\lambda/\mu}=f_{\hat  \mu'/\hat \lambda'}$ concludes the proof.

\end{document}